\documentclass[12pt]{article}

\usepackage{amsmath}
\usepackage{graphicx}
\usepackage{amssymb}
\usepackage{amsfonts}
\usepackage{latexsym}

\DeclareMathOperator{\cx}{\square}

\def\beq{\begin{eqnarray}}
\def\eeq{\end{eqnarray}}

\def\al{\alpha}
\def\be{\beta}
\def\ch{\chi}
\def\ga{\gamma}
\def\de{\delta}
\def\vp{\varepsilon}

\def\ze{\zeta}

\def\ka{\kappa}
\def\la{\lambda}
\def\na{\nabla}

\def\si{\sigma}
\def\om{\omega}
\def\ph{\varphi}

\def\De{\Delta}
\def\La{\Lambda}

\begin{document}
\begin{center}

{\large\bf On the conformal properties of topological terms
\\
in even dimensions}
\vskip 4mm

Fabr\'{\i}cio M. Ferreira~$^{a,c}$,
\
Ilya L. Shapiro~$^{a,b}$
\
and
\
Poliane M. Teixeira~$^{a}$

\end{center}
\vskip 4mm
\begin{center}
{\sl
(a) \ Departamento de F\'{\i}sica, ICE, Universidade Federal de Juiz de Fora
\\
Campus Universit\'{a}rio - Juiz de Fora, 36036-330, MG, Brazil
\\
E-mail addresses: fabricio.ferreira@ifsudestemg.edu.br, \ \
shapiro@fisica.ufjf.br, \ \
poliane@fisica.ufjf.br

\vskip 2mm

(b) \ Tomsk State Pedagogical University and Tomsk State
University, Tomsk, Russia
\vskip 2mm

(c) \ Instituto Federal de Educa\c c\~ao, Ci\^encia e Tecnologia
do Sudeste de Minas Gerais \\
IF Sudeste MG - Juiz de Fora, 36080-001, MG, Brazil
\vskip 2mm
}
\end{center}
\vskip 6mm

\begin{quotation}
\noindent
\textbf{Abstract.} \
Conformal properties of the topological gravitational terms in
$D=2$, $D=4$ and $D=6$ are discussed. It is shown that in the last
two cases the integrands of these terms, when being settled into
the dimension $D-1$ and multiplied by a scalar, become conformal
invariant. Furthermore we present a simple covariant derivation
of Paneitz operator in $D=4$ and formulate two general conjectures
concerning the conformal properties of topological structures in
even dimensions.
\vskip 3mm

\vskip 2mm

\vskip 2mm

\end{quotation}

\section{Introduction}
\label{intro}

Conformal operators and conformal properties of topological terms in
different space-time dimensions $D$ are important issues, especially
due to the applications in quantum theory. The solution in $D=2$ is
a very well-known Polyakov action \cite{Polyakov}, while the
conformal operator is just a two-derivative $\Delta_2=\Box$.
The conformal operator $\De_4$
acting on scalar in $D=4$ was first obtained by Paneitz \cite{Paneitz}
and independently by Riegers, Fradkin and Tseytlin \cite{rei}. This
operator has four derivatives and acts on the conformally inert
scalar field. One can easily obtain a generalization to the   $D\neq 4$,
when the corresponding scalar gains a non-trivial transformation law
proportional to the difference $D-4$, see Ref. \cite{BarrosShapiro}.

In order to integrate conformal anomaly in $D \geq 6$ and explore the
allegedly general universality properties, it would be very  useful to
have similar operators in general even dimensions $D=6,8,...$.
In the mathematical literature one can find a general theory for
constructing conformal operators \cite{Conf1,Conf2,Conf3,Branson-85},
which can be used to obtain explicit examples. For instance, the
analog of Paneitz operator with six derivatives, $\De_6$,  in $D=6$
can be found in \cite{Arak}, consequent paper \cite{Hamada} and in
\cite{Osborn}, where  the generalization to  $D\neq 6$ was also obtained.

It is important to remember that the generalization of
the results of \cite{Polyakov,rei} to dimensions $D \geq 6$ requires
not only constructing the corresponding conformal operators, but also
their relations to the topological terms\footnote{This was recently
discussed in \cite{Ansel}, where one the our conjectures from Sect. 4
was formulated for the particular case of a flat background. Here we
approach the problem in a partially different way and consider an
arbitrary curved metric. }. By using both things and also the
relation between surface terms in the anomaly and local finite terms
in the effective action, one can expect to obtain compact and useful
expressions for the anomaly-induced effective action of gravity, such
as Polyakov action in $D=2$ and analogous expression in $D=4$
\cite{rei} (see also \cite{PoImpo} and \cite{MaMo} for the reviews and
description of recent results in this direction).

In the present work we describe some preliminary results related
to the conformal properties of topological structures in even
dimensions and their relation to the conformal invariants in odd
dimensions. Furthermore, we formulate two conjectures about
possible relation between the integrands of topological structures
and existence of higher derivative conformal operators, which may
be valid (or not) in even dimensions. The verification of these
conjectures will be hopefully presented soon in a separate paper.
The material presented here is very simple and some part of it may
be not completely new. However we believe that it may have some
interest for those who work on the related subjects. In order to
make the manuscript brief we skip many details
concerning conformal transformations of curvature tensor(s) and
their contractions. One can consult the previous paper \cite{Stud}
for all intermediate formulas. At the same time, all relevant final
expressions are presented for the convenience of the reader.

The paper is organized as follows. In Sect. 2 one can find some
covariant calculations, which includes a new way of deriving the
Paneitz operator in $D=4$. Sect. 3 is devoted to the conformal
transformation of the integrands of topological invariants in
$D=2$, $D=4$ and $D=6$ dimensions. As a by-product we arrive at
the new conformal invariants in $D-1$  dimensions for all three
cases. In Sect. 4 the two conjectures about conformal operators
and conformal properties of topological structures are formulated.
Finally, in Sect. 5 we draw our conclusions.

\section{Covariant derivation of Paneitz operator}
\label{sec:1}

Let us start by reviewing terms which are topological in $D=2$
(Einstein-Hilbert) and $D=4$ (Gauss-Bonnet term). Previously, the
last case has been discussed in some works devoted to quantum
gravity \cite{capperkimber,Weyl}, where one can find more
detailed consideration.

In $D=2$ one has to start from the Einstein-Hilbert action
\beq
\label{GB1}
S_2 &=& \int d^Dx \sqrt{-g}\,R \,,\qquad D=2\,.
\eeq
The equations of motion boil down to
\beq
\label{GB3}
R^{\mu\nu} - \frac{1}{2}\,g^{\mu\nu}\,R = 0\,,
\eeq
which is an identity in $D=2$.

It is interesting to see whether something similar occurs in $D=4$.
In this case the topological action has the form
\beq
\label{GB4}
S_4 &=& \int d^4x \sqrt{-g} \ E_4 \ ,
\eeq
where
$\,E_4 = R^{\al\be\rho\si}R_{\al\be\rho\si} - 4 R^{\mu\nu}R_{\mu\nu} + R^2$
is Euler characteristic in $D=4$.

It proves useful to define the integrals of the squares of curvatures,
\beq
\label{GB5}
I_1(D) = \int d^Dx \sqrt{-g} \,R^2_{\mu\nu\alpha\beta} \,,
\quad
I_2(D) = \int d^Dx \sqrt{-g} \,R^2_{\mu\nu} \,,
\quad
I_3(D) = \int d^Dx \sqrt{-g} \,R^2\quad
\eeq
and the surface term $\,I_4(D) = \int d^Dx \sqrt{-g} \,{\cx} R$. \
The variation with respect to the metric in $D=4$ yields
\beq
\label{GB7}
\nonumber
\frac{1}{\sqrt{-g}}\frac{\delta I_1(4)}{\delta g_{\mu\nu}}
&=&
\frac{1}{2}g^{\mu\nu}R^2_{\rho\sigma\alpha\beta}
- 2R^{\mu\sigma\alpha\beta}R_{\sigma\alpha\beta}^{\nu}
- 4R^{\mu\alpha\nu\beta}R_{\alpha\beta} + 4R^{\mu}_{\alpha}R^{\nu\alpha}
\\
&+& 2\nabla^{\mu}\nabla^{\nu}R - 4{\cx} R^{\mu\nu} \,,
\\
\label{GB8}
\frac{1}{\sqrt{-g}}\frac{\delta I_2(4)}{\delta g_{\mu\nu}}
&=&
\frac{1}{2}g^{\mu\nu}R^2_{\rho\si}
- 2R^{\mu\alpha\nu\beta}R_{\al\be} + \nabla^{\mu}\nabla^{\nu}R
- \frac{1}{2}g^{\mu\nu}{\cx} R - {\cx} R^{\mu\nu} \,,
\\
\label{GB9}
\frac{1}{\sqrt{-g}}\frac{\delta I_3(4)}{\delta g_{\mu\nu}}
&=&
\frac{1}{2}g^{\mu\nu}R^2 + 2\nabla^{\mu}\nabla^{\nu}R
- 2g^{\mu\nu}{\cx} R - 2RR^{\mu\nu} \,.
\eeq
It is not easy to show that the linear combination of these
expressions corresponding to the action (\ref{GB4}) is identically
zero, as it was discussed in \cite{capperkimber}. At the same
time the traces of the combinations corresponding to the Weyl
action $I_1-2I_2+I_3/3$ and to the Gauss-Bonnet topological
term (\ref{GB4}) can be immediately observed to vanish.

For $D\neq 4$ the Gauss-Bonnet term (\ref{GB4}) is not topological.
It is easy to derive the trace of the corresponding equation,
\beq
\label{GB15}
\frac{1}{\sqrt{-g}}g_{\mu\nu}\frac{\delta}{\delta g_{\mu\nu}}
\int d^D x \sqrt{-g} \,E_4
&=&
\frac{(D-4)}{2}\,E_4 \,.
\eeq

Consider equations of motion for the action
\beq
\label{GB16}
I_E(D) &=& I_1(D) - 4I_2(D) + I_3(D)
\eeq
on a special de~Sitter background, when Riemann and
Ricci tensors can be presented as
\beq
\label{GB17}
R_{\mu\nu\al\be}
&=& \frac{1}{D(D-1)}\,\La\,(g_{\mu\al}g_{\nu\be} - g_{\mu\be}g_{\nu\al})
\,, \quad
R_{\mu\nu} = \frac{1}{D}\,\La\,g_{\mu\nu}
\,,\quad
\La=const \,.
\eeq

After a small algebra we arrive at the following results:
\beq
\label{GB18}
\frac{1}{\sqrt{-g}}\frac{\delta I_1}{\delta g_{\mu\nu}}\bigg|_{dS}
= \frac{(D-4)}{D^2(D-1)} \ {\Lambda}^2g^{\mu\nu} \,,
\eeq
\beq\label{GB19}
\frac{1}{\sqrt{-g}}\frac{\delta I_2}{\delta g_{\mu\nu}}\bigg|_{dS}
= \frac{(D-4)}{2 D^2} \ {\Lambda}^2g^{\mu\nu} \,,
\eeq
\beq\label{GB20}
\frac{1}{\sqrt{-g}}\frac{\delta I_3}{\delta g_{\mu\nu}}\bigg|_{dS}
= \frac{(D-4)}{2 D} \ {\Lambda}^2g^{\mu\nu}\,.
\eeq
Consequently, for the $D$-dimensional version of the Gauss-Bonnet
term we obtain
\beq
\label{GB21}
\frac{1}{\sqrt{-g}}
\frac{\delta I_E(D)}{\delta g_{\mu\nu}} \bigg|_{dS}
&=& \frac{(D-2)(D-3)(D-4)}{2D^2(D-1)}\,{\Lambda}^2g^{\mu\nu} \, .
\eeq
Naturally, when $D=4$, the equation (\ref{GB21}) becomes zero, but the
same also occurs in $D=2$ and $D=3$ cases, where the Gauss-Bonnet term
is not topological. One can note that the expressions (\ref{GB18}) do
vanish only at $D=4$, so the cancelation for $D=2$ and $D=3$ takes
place only for the topological term. Later on we obtain more detailed
form of this relation.

The next exercise is to obtain the Paneitz operator \cite{Paneitz}
in $D=4$ in a covariant way. The usual definition of this operator
is through the conformal transformation,
\beq
\label{conform}
g_{\mu\nu}
&=&
{\bar g}_{\mu\nu}\,e^{2\si}\,,\quad
\mbox{where} \quad
\si=\si(x)
\eeq
and ${\bar g}_{\mu\nu}$ is a fiducial metric. We can assume that
$\si$ is a scalar field and then ${\bar g}_{\mu\nu}$ is a second
rank tensor. It is important that ${\bar g}_{\mu\nu}$ does not
depend on $\si$ and this can be achieved, e.g., by using the
covariant non-local construction of \cite{ETG-76,FrVi-78}.

For the conformally inert scalar $\ph={\bar \ph}$ the Paneitz
operator $\De_4$ has to be Hermitian and provide the invariance
of the bilinear expression,
\beq
\int d^4x\sqrt{-g}\,\ph \De_4 \ph
&=&
\int d^4x\sqrt{-{\bar g}}\,{\bar \ph} {\bar \De}_4 {\bar \ph}\,.
\label{Pan}
\eeq
Here the bar means that the operator is constructed with the
${\bar g}_{\mu\nu}$ metric. The solution for $\De_4$ has been
found in \cite{Paneitz} (see also \cite{rei} and generalization
to other dimensions in \cite{BarrosShapiro}), but we shall
solve the same problem in a completely covariant way, without
explicit use of the transformation (\ref{conform}).

We start from a simple observation about the variational
derivative with respect to $\si$ in (\ref{conform}). For a
functional of a metric $A=A(g_{\mu\nu})$ we have
\beq
\frac{\delta A}{\delta \si}
&=&
\frac{\delta g_{\mu\nu}}{\delta \si}
\,\cdot\,
\frac{\delta A}{\delta g_{\mu\nu}}
\,=\,
2\,{\bar g}_{\mu\nu}\,e^{2\si}\,
\frac{\delta A}{\delta g_{\mu\nu}}
\,=\,
2\,g_{\mu\nu}\,\frac{\delta A}{\delta g_{\mu\nu}}\,.
\label{ident}
\eeq
This simple calculation shows that everything that is linear in
$\si$ can be obtained by taking the trace of covariant equations
of motion for the metric.

In order to obtain the Paneitz operator $\Delta_4$ in a covariant
way, one can define new actions which depend on an additional
conformally inert scalar field $\,\ph={\bar \ph}$,
\beq
\label{GB22}
I^\ph_1 = \int d^4x \sqrt{-g}\,\ph R^2 \,,
\quad
I^\ph_2 = \int d^4x \sqrt{-g}\,\ph R^2_{\mu\nu} \,,
\\
I^\ph_3 = \int d^4x \sqrt{-g}\,\ph R^2_{\mu\nu\alpha\beta}
\,,
\quad
I^\ph_4 = \int d^4x \sqrt{-g}\,\ph {\cx} R \,.
\eeq

The equations of motion have the form
\beq
\label{GB23}
\frac{1}{\sqrt{-g}}\frac{\delta I^\ph_1}{\delta g_{\mu\nu}}
&=&
\frac{1}{2}g^{\mu\nu}R^2{\ph}
+ 2\nabla^{\nu}\nabla^{\mu}(R\varphi)
- 2g^{\mu\nu}{\cx}(R\varphi)
- 2R^{\mu\nu}(R\varphi) \,,
\\
\nonumber
\frac{1}{\sqrt{-g}}\frac{\delta I^\ph_2}{\delta g_{\mu\nu}}
&=&
\frac{1}{2}g^{\mu\nu}R^2_{\rho\si}{\varphi}
- 2R^{\nu\alpha}R^{\mu}_{\al}{\varphi}
+ 2\nabla_{\lambda}\nabla_{\mu}(R^{\lambda}_{\nu}{\varphi})
\\
\label{GB24}
&-& g^{\mu\nu}\nabla_{\beta}\nabla_{\alpha}(R^{\alpha\beta}{\varphi})
- {\cx}(R^{\mu\nu}{\varphi}) \,,
\\
\label{GB25}
\frac{1}{\sqrt{-g}}\frac{\delta I^\ph_3}{\delta g_{\mu\nu}}
&=&
\frac{1}{2}g^{\mu\nu}R^2_{\alpha\beta\rho\sigma}{\varphi}
+ 4\nabla_{\beta}\nabla_{\alpha}(R^{\mu\beta\alpha\nu}{\varphi})
- 2(R^{\mu}_{\cdot\beta\alpha\rho}R^{\nu\beta\alpha\rho}{\varphi}) \,,
\\
\nonumber
\frac{1}{\sqrt{-g}}\frac{\delta I^\ph_4}{\delta g_{\mu\nu}}
&=& \frac{1}{2}g^{\mu\nu}R \cx{\varphi}
+ \nabla^{\mu}\nabla^{\nu}\cx{\varphi}
- g^{\mu\nu}{\cx}^2{\varphi}
- R^{\mu\nu}{\cx} {\varphi}
- R\nabla^{\mu}\nabla^{\nu}{\varphi}
\\
\label{GB26}
&+& \nabla_{\lambda}(Rg^{\lambda(\mu}\nabla^{\nu)}{\varphi})
- \frac{1}{2}\nabla_{\lambda}(g^{\mu\nu}R\nabla^{\lambda}{\varphi}) \,.
\eeq

Using this formulas it is easy to check that the Weyl term with an
additional scalar is still conformal invariant,
\beq
\label{GB31}
\nonumber
\frac{1}{\sqrt{-g}}\,g_{\mu\nu}\,\frac{\delta}{\delta g_{\mu\nu}}
\int d^4x \sqrt{-g}\,\ph C^2
&=& 0 \,.
\eeq
For the Gauss-Bonnet term with additional scalar we obtain
\beq
\label{GB32}
\frac{1}{\sqrt{-g}}\,g_{\mu\nu}\frac{\delta}{\delta g_{\mu\nu}}
\int d^4x \sqrt{-g}\,\ph\,E_4
&=&
4R^{\mu\nu}(\na_{\mu}\na_{\nu}\ph) \,-\, 2R {\cx} \ph \,.
\eeq

Let us now make an important step and introduce the ``corrected''
term
\beq
{\tilde E}_4 &=& E_4 - \frac23\,{\cx} R\,.
\label{GBcorr}
\eeq
Taking into account (\ref{GB32}) and (\ref{GB26}), after a
very small algebra we arrive at
\beq
\nonumber
&& \frac{1}{2\sqrt{-g}}\,g_{\mu\nu}\,\frac{\delta}{\delta g_{\mu\nu}}
\int d^4x \sqrt{-g}\,\ph\,{\tilde E}_4
\\
\label{GB33}
&&
=\,
\Big[{\cx}^2 + 2R^{\mu\nu}\nabla_{\mu}\nabla_{\nu} - \frac{2}{3}R \cx + \frac{1}{3}(\nabla_{\lambda}R)\nabla^{\lambda}\Big] \ph
\,=\,\De_4\ph \,,
\eeq
where $\De_4$ is exactly the Paneitz operator which we are looking for.

Two observations are in order. First, the derivation of
the conformal operator $\De_2=\cx$ in $D=2$ can be performed in the
very same way as we just did in $D=4$, but in the two-dimensional case
there is no need to introduce an additional term to $E_2=R$. Since the
calculation is quite trivial, we skip the details.
The second point is that there is {\it no regular way} to derive
the coefficient in Eq. (\ref{GBcorr}), so the origin of its value
$-2/3$ looks
mysterious. In the full conformal derivation (see, e.g., \cite{Stud})
this coefficient provides cancellation of all but linear terms in
$\si$ in the transformation of  $\sqrt{-g}{\tilde E}_4$, but (as
far as we know) there is no other way to obtain this coefficient
except by an explicit calculation.
\section {Conformal transformation of topological terms}
\label{sec:2}

Let us explore the conformal properties of topological terms in even
dimensions. We will be mainly concerned with the $D=6$ case which
attracted considerable interest in the recent literature (see, e.g.,
\cite{Osborn,Grinshtein,Solodukhin,Intr} and references therein).
According to the standard classification \cite{DeserSchwimmer}
(see also earlier work \cite{DDI-76}), the
anomalous terms that correspond to the non-local part of effective
action are either conformal invariants or topological terms. Hence
it is very important to better understand the conformal properties
of the topological terms, in particular for the case of $D=6$.

The conformal transformation is defined as a parametrization
(\ref{conform}) of the metric tensor. It makes sense to analyze
the transformations of Euler densities not only in the dimensions
in which these quantities are topological, but also in other
dimensions. Euler density in even dimensions $D=2n$ ($n = 1, 2, ...$)
is well-known (see e.g. \cite{decanini}),
\beq
\label{6d01}
E_{2n}
&=& \frac{1}{2^{n}}\,
\vp^{\al_1\be_1\,\dots\,\al_n \be_n} \vp^{\ga_1 \delta_1\,\dots\,\ga_n \delta_n}\,
 R_{\al_{1} \be_{1} \ga_{1} \delta_{1}}
 \,\,\,\dots\,\,\,
 R_{\al_{n} \be_{n} \ga_{n} \delta_{n}}\,.
\eeq

It is instructive to consider a few examples.  For $D=2$,
the definition (\ref{6d01}) gives
\beq 	
\label{6d02}
E_2
&=&
\frac{1}{2}\,\vp^{\mu \nu} \vp^{\al \be}
R_{\mu \nu \al \be} = R\,.
\eeq
In $D=4$ case Eq. (\ref{6d01}) provides the Gauss-Bonnet term (\ref{GB4}),
\beq
\label{6d03}
E_4
&=&
\frac{1}{4}\,\vp^{\mu \nu \la \tau} \vp^{\al \be \rho \si}
\,R_{\mu \nu \al \be}R_{\la \tau \rho\si}\, .
\eeq
In $D=6$ the evaluation of (\ref{6d01}) is more cumbersome,
the result is (see, e.g., \cite{Myers})
\beq
\nonumber
E_6
&=& \frac{1}{8}\,\vp^{\mu \nu \al \be \la \xi}\,
\vp^{\rho \si \ka \om \eta \ch }\,
R_{\mu \nu \rho \si}\,R_{\al \be \ka \om}\,R_{\la \xi \eta \ch}
\\
\nonumber
&=&
4R^{\mu \nu}\,_{\al \be} {R}^{\al \be}\,_{\la \tau}
R^{\la \tau}\,_{\mu \nu}
- 8 {R}_{\mu\al\nu\be}{R}^{\al}\,_{\la}\,^{\be}\,_{\tau}
{R}^{\la\mu\tau\nu}\,
-24R^{\nu}_{\mu}R_{\al \be \la \nu}R^{\al \be \la \mu }
\\
&+&
24R_{\mu \al \nu \be }R^{\mu \nu}R^{\al \be}\,
+16R^{\nu}_{\mu}R^{\mu}_{\al}R^{\al}_{\nu}
+3RR^{2}_{\mu \nu \al \be}
-12RR^{2}_{\mu \nu}+R^{3}\,.
\label{6d04}
\eeq

Consider how these three quantities behave under $D$-dimensional
conformal transformation (\ref{conform}). The transformations of
Riemann, Ricci tensor and scalar and of the $\Box$ can be found,
e.g., in \cite{Stud}, so let us skip the intermediate formulas
and present only the final results,
\beq
\label{6d07}
\sqrt{-g}\,E_2
&=&
\sqrt{-\bar{g}}\,e^{(D-2)\si}\,
\Big\{\bar{E}_2-(D-1)\Big[ 2\bar{\cx}\si
+(D-2)(\bar{\nabla}\si)^2 \Big]\Big\},
\eeq
where we multiplied the expression by $\sqrt{-g}$ for convenience.

Similarly, the $E_4$ calculation yields
\beq
\label{6d08}
\sqrt{-g}E_4
&=&
\sqrt{-\bar{g}}\, e^{(D-4)\si}\,
\big\{ \bar{E}_4 + (D-3)\chi_{(4)}\big\}\,,
\eeq
where
\beq
\chi_{(4)}
&=&
8\bar{R}_{\mu \nu}\si^{\mu\nu}
- 8 \bar{R}_{\mu \nu}\si^\mu \si^\nu
- 4\bar{R}{\bar{\cx}}{\si}
- 2 (D-4)\bar{R}(\bar{\nabla}{\si})^{2}
\,+\,
(D-2)\big[8\si_{\mu\nu}\si^\mu \si^\nu
\nonumber
\\
&-&
4 \si_{\mu\nu}^2
+ (D-4)(D-1)(\bar{\nabla}{\si})^{4}
+ 4(\bar{\cx}{\si})^2
+ 4(D-3)\bar{\cx}{\si} (\bar{\nabla}{\si})^2
\big]
\,.
\eeq
Here we used the condensed notations
$\,\si_\al=\bar{\na}_\al\si$,
$\,\si_{\al\be}=\bar{\na}_\al\bar{\na}_\be \si$,
$\,\bar{\cx}={\bar g}^{\al\be}\si_{\al\be}$,
$\,(\bar{\nabla}\si)^2={\bar g}^{\al\be}\si_\al\si_\be$,
also all indices are raised and lowered with the fiducial
metric ${\bar g}^{\al\be}$ and its inverse.

One can note that the conformally non-covariant part of the
{\it r.h.s.} of (\ref{6d07}) is proportional to $D-3$, which
is a non-linear generalization of the previously considered
Eq. (\ref{GB21}). One can see that the non-linear expression
(\ref{6d08}) is conformally non-covariant at $D=2$ (some
related observations can be found in \cite{Stud}), but is
covariant at $D=3$.

Finally, consider the case of $E_6$. The corresponding calculations
were performed by using {\sl Cadabra} software \cite{cadabra01,cadabra02}
and the result reads
\beq
\label{6d09}
\sqrt{-g}E_{6}
&=&
\sqrt{-\bar{g}}\,e^{(D-6)\si}
\,\big\{
\bar{E}_6
+ (D-5)\,\chi_{(6)} \big\}\,,
\eeq
where
\beq
\chi_{(6)}
&=&
- \,\big[
6\bar{\cx}\si \,+\, 3(D-6)  (\bar{\nabla}\si)^2\big]\bar{E}_4
\nonumber
\\
&+&
24 \big(
2 \bar{R}_{\al\be}\bar{R}^{\mu\al\nu\be}
-\bar{R}_{\al\be\ga}\,^\nu\bar{R}^{\al\be\ga\mu}
- \bar{R}\bar{R}^{\mu\nu}
+ 2 \bar{R}^{\mu\al}\bar{R}^\nu_\al
\big)\big(\si_{\mu}\si_{\nu}- \si_{\mu\nu}\big)
\nonumber
\\
&+&
24 (D-4)\bar{R}^{\mu\al\nu\be} \si_{\mu\nu}
\big(\si_{\al\be}
- 2 \si_\al \si_\be\big)
+ 48 (D-4)\bar{R}^{\mu}_{\nu}\big(\si_{\mu\al}\si^{\nu\al}
- \si_\mu^\nu \bar{\cx}\si\big)
\nonumber
\\
&+&
48(D-4)\bar{R}^{\mu\nu}\big( \si_\mu \si_\nu \bar{\cx}\si
- 2\si_{\mu\al} \si_\nu^\al\big)
+ 12 (D-4)\bar{R} \big[(\bar{\cx}\si)^{2}
- \si_{\mu\nu}^2
+ 2\si_{\mu\nu}\si^\mu \si^\nu\big]
\nonumber
\\
&-& 24(D-4)\bar{R}^{\mu \nu}(\bar{\nabla}{\si})^2
\big[(D-5)\si_{\mu\nu}
- (D-3) \si_{\mu}\si_\nu\big]
\nonumber
\\
&+&
12(D-5)(D-4)\bar{R}\bar{\cx}\si(\bar{\nabla}\si)^2
+ 3(D-6)(D-4)(D-3)\bar{R}(\bar{\nabla}{\si})^{4}
\nonumber
\\
&+&
  8 (D-4)(D-3)\big[3 \si_{\mu\nu}^2\,\bar{\cx}\si
- 2\si_\mu^\nu\,\si_\nu^\al\,\si_\al^\mu
+ 6 \si_\mu^\nu\,\si_\nu^\al\,\si_\al \si^\mu
- (\bar{\cx}\si )^3\big]
\nonumber
\\
&+&
12(D-4)^2 (D-3)\,(\bar{\nabla}{\si})^2\,
\big[\si_{\mu\nu}^2 - (\bar{\cx}\si)^2\big]
\label{big chi}
\\
&-&
24(D-4)(D-3)\,\si_{\mu\nu}\,\si^\mu \si^\nu
\big[(D-2)(\bar{\nabla}{\si})^2
+ 2\bar{\cx}\si\big]
\nonumber
\\
&-&
(D-4)(D-3)(D-2)\big[
6(D-5)\bar{\cx}\si
\,+\,
(D-6)(D-1)(\bar{\nabla}{\si})^2\big](\bar{\nabla}\si)^4 \,.
\nonumber
\eeq

An interesting feature of Eq. (\ref{6d09}) is that the conformally
non-covariant part of this expression vanish in $D=5$ dimension.
As we have seen before, this is similar to $E_2$ and $E_4$. As a
consequence one can construct new conformal invariants in odd
dimensions $2n-1$. Consider an auxiliary scalar field $\Phi$
which transforms according to
\beq
\label{Phi trans}
\Phi &=& e^{\si}\,{\bar \Phi}
\eeq
simultaneously with (\ref{conform}). Then we meet
\beq
\label{D-1 invs}
\int d^{2n-1}x\sqrt{-g}\,\Phi\,E_{2n}
&=&
\int d^{2n-1}x\sqrt{-{\bar g}}\,{\bar \Phi}\,{\bar E}_{2n}\,,
\eeq
where $n=1,2,3$ and the expressions (\ref{D-1 invs}) provides the
set of conformally invariant actions. Of course this is a trivial
statement for $D=2$, but in the cases of $D=4$ and $D=6$ we can
claim that the topological invariants in these even dimensions
give rise to the new conformal invariants (\ref{D-1 invs}) in
three- and five-dimensional spaces, correspondingly.
\section{Conjectures}
\label{SectConj}

Taking our previous experiences into account, let can formulate
the following two conjectures concerning the topological terms
(\ref{6d01}):
\vskip 3mm

{\bf Conjecture 1.} \
For any even dimension $D=2n,\,n=1,2,3,4,...$, \
the expressions (\ref{D-1 invs}) are conformal invariant
if the scalar $\Phi$ transforms according to (\ref{Phi trans}).
This means we arrive at the chain of conformal actions
\beq
S^c_{2n-1} &=& \int d^{2n-1}x \sqrt{-g}\,\Phi\,E_{2n}
\label{2n-1}
\eeq
in odd dimensions.
\vskip 3mm

{\bf Conjecture 2.} \
For any even dimension $D=2n$ there is such a metric-dependent
vector function $\chi^\mu_{2n}$ that the ``corrected'' topological
invariant
\beq
E_{2n}+\na_\mu \chi^\mu_{2n}\,,
\label{Cortop}
\eeq
possesses linear conformal transformation,
\beq
\sqrt{-g}\big(E_{2n}+\na_\mu \chi^\mu_{2n}\big)
&=&
\sqrt{-{\bar g}}\big({\bar E}_{2n}
+ {\bar \na}_\mu {\bar \chi}^\mu_{2n}
+ c\cdot {\bar \De}_{2n}\si\big)\,.
\label{conf top}
\eeq
Here $c$ is some unknown constant and operator
$\,\De_{2n}=\Box^n\,+\,\dots\,$ is conformal, in the
sense
\beq
\int d^{2n}\sqrt{-g}\,\ph \De_{2n} \ph
&=&
\int d^{2n}\sqrt{-{\bar g}}\,\ph {\bar \De}_{2n} \ph\,.
\label{Pan D}
\eeq
Let us remember that all quantities with bars are constructed on the
basis of the fiducial metric ${\bar g}_{\mu\nu}$ in (\ref{conform}).
In the case of $D=2$ we know that $\chi^\mu_2=0$ and for $D=4$ we
know that $\chi^\mu_4=-(2/3)\na^\mu R$. The verification of this
conjecture for six dimensions requires a significant calculational
work and we expect to report on the result
soon\footnote{After submitting the first version of this work
we learned that the flat limit of the relation (\ref{conf top})
and an incomplete form of anomaly-induced action was recently
obtained in a very interesting paper \cite{Elvang}. Our project
can be seen as presenting the result in a covariant form
of relation (\ref{conf top}).} \cite{FaSh}.

An important step towards a general understanding of the second
Conjecture would be explanation of the $-2/3$ coefficient in the
four-dimensional case. At the moment we are not able to give
such an explanation and rely on a direct computations.
\section{Conclusions and discussions}
\label{concl}

Since the conformal anomaly is one of the main sources of our
knowledge of the semiclassical corrections to the gravitational
action (see, e.g., \cite{PoImpo,birdav,duff94,Deser-N}), it would
be useful to have better understanding of the conformal properties
of the terms which constitute this anomaly. In this relation it is a
challenging problem to establish conformal properties of the
topological invariants and their relations to the conformal
operators acting on conformally inert scalars.

At the moment we know such relations for the two- and
four-dimensional spaces. However, there is no real understanding
of the fundamental reasons of why these relations take place in
$D=4$ and whether similar relations exist for higher even dimensions.
In this respect it would be most interesting to verify the second
Conjecture formulated above (see also previous work \cite{Ansel} on
the same subject). A practical realization of this program is a
necessary step in  integrating conformal anomaly in $D=6$ and
higher even dimensions, and also may help to approaching
the solution of one of the mathematical puzzles related to
conformal anomaly. It is important  that integrating the trace anomaly
requires not only conformal operator \cite{Branson-85,Osborn}
(see also \cite{Conf3,Wunsch-86,GoverPeterson,Waldron} for
other publications on the subject), but also the relation between
conformal operators and topological structures, e.g., expressed
in the form (\ref{conf top}). This kind of formula is critically
important for integrating anomaly in $D=2$ and $D=4$ and hence
the proof of the Conjecture 2 would be a decisive step forward
in completing the same program in higher even dimensions.

After the proof of the second Conjecture, the problem will not be
solved yet. The reason is that there a third type of terms in the
anomaly, which go beyond the known classification of
\cite{DeserSchwimmer} and come from the renormalization of
surface terms. The experience which we have from the $D=4$
shows that these terms should be taken seriously, in particular
they emerge in a direct calculation via adiabatic regularization
(see, e.g., \cite{ParkerToms}). In order to neglect these terms
one needs at least to be sure that the finite anomaly-generating
terms in the effective action are local. The detailed constructive
proof of this statement would open the door for deriving
anomaly-induced action in  $D=6,8,...$ and to the corresponding
physical and mathematical applications.

\section*{Appendix. \ The case of conformally flat metric}
\hspace{0.45cm} In order to show how the Conjecture 2 described in Sect.
\ref{SectConj} works, let us consider a relatively simple example
of the conformally flat metric. Earlier the same method has been
used in \cite{Ansel}, but we shall go into more details and obtain
a slightly more general result.

Let us remember that our ansatz assumes adding only total second
derivatives terms to $E_6$. Using some reduction procedure, one
can show that the general form  of the corrected topological term is
\beq
\tilde{E}_6 &=&
E_6 + \al_1{\cx}^{2}R
+\al_2{\cx}R^2+\al_3{\cx}R^2_{\mu\nu}+
\al_4{\cx}R^2_{\mu\nu\al\be}
+\al_5\nabla_{\mu}\nabla_{\nu}
(R^{\mu}\,_{\la\al\be}R^{\nu\la\al\be})
\nonumber
\\
&+&
\al_6  \nabla_{\mu}\nabla_{\nu}(R_{\al\be}R^{\mu\al\nu\be})
+ \al_7 \nabla_{\mu}\nabla_{\nu}(R^\mu_\la R^{\nu \la}) + \al_8
\nabla_{\mu}\nabla_{\nu}(RR^{\mu\nu})\,.
\label{EGeral}
\eeq
Our interest is to find the values of the parameters $\al_{1,2,...,8}$
for which the relation (\ref{conf top}) takes place. In order to do it,
we can choose any background metric, which makes the solution
simpler. One of the possibilities is to take the  metric
 $\,g_{\mu\nu}= e^{2\si}\eta_{\mu\nu}$, with $\,\si=\si(\eta)$,
where $\eta$ is conformal time. After that one has to evaluate the
conformal transformation for each of the nine terms of  (\ref{EGeral}).

After performing these  steps we found the combination of $\al's$ which
satisfies the requested condition. The most general form depends on four
parameters $a,b,c,d$ as follows:
\beq
\label{E6_tilde02}
\tilde{E}_6 &=&
E_6+\frac{3}{5}{\cx}^{2}R
+a{\cx}R^2+b{\cx}R^2_{\mu\nu}+c{\cx}R^2_{\mu\nu\al\be}
+d\nabla_{\mu}\nabla_{\nu}(R^{\mu}\,_{\la\al\be}R^{\nu\la\al\be})
\nonumber
\\
&-& \Big(\frac{21}{5}+20a+8b+6c+\frac{d}{2} \Big) \nabla_{\mu}\nabla_{\nu}(R_{\al\be}R^{\mu\al\nu\be})
+ \Big(3-2b-2c-\frac{d}{2} \Big)
\nabla_{\mu}\nabla_{\nu}(R^\mu_\la R^{\nu \la})
\nonumber
\\
&+&
\frac{1}{5}\Big(-3+6b+6c+\frac{d}{2}
\Big)
\nabla_{\mu}\nabla_{\nu}(RR^{\mu\nu})\,.
\label{EGeral-1}
\eeq
This particular form of (\ref{EGeral}) eliminates all  terms of second
and higher orders in $\si$ in a conformal transformation of the
action (\ref{E6_tilde02}).  Let us note that the relation
(\ref{E6_tilde02}) possesses this property for any  $a, b, c, d$.

Indeed, the cancelation of the second- and higher-orders in
$\si$ on flat background does not guarantee that the desired
relation  (\ref{conf top}) holds on an arbitrary background.
One can suppose that this relation on an arbitrary background will
require to impose some constraints on the parameters  $a, b, c, d$.
Anyway, in order to have a chance to achieve the result, it is
certainly important to have a most general form, such as
(\ref{EGeral-1}).

Let us consider some particular cases of (\ref{EGeral-1}). For instance,
one can take the parameters $a, b, c$ and $d$ in such a way that the
terms with Riemann tensor vanish, that means one-parametric
solution,
\beq
c=d=0\,,\quad
a=\frac{123}{100}-\frac{1}{2}\ze
\,,\quad
b=\frac{5}{4}\ze-\frac{18}{15}\,,
\label{An}
\eeq
that corresponds to the result of \cite{Ansel}. Another, much simpler
case is obviously $a=b=c=d=0$, corresponding to
\beq
\sqrt{-g}\Big\{E_6 +\frac{3}{5}\big[
{\cx^{2}}{R}
+\nabla_{\mu}\nabla_{\nu}( -7R_{\al\be}R^{\mu\al\nu\be}
+5R^{\mu}_{\al}R^{\nu\al}
-RR^{\mu\nu})
\big]
\Big\}=-6\Delta_6\si
\eeq
on a flat metric background.

The next phase of the work includes trying to find similar relations
for an arbitrary fiducial metric. In this way one can expect to remove
the remaining uncertainty of the coefficients $a, b ,c$ and $d$ in (\ref{E6_tilde02}). In this way one can establish a general form of
$\Delta_6$ which can be used for comparison with
\cite{Arak,Hamada,Osborn}, verification of Conjecture 2 and
finally for deriving the anomaly-induced effective action in $D=6$.

\section*{Acknowledgements}

I. Sh. is grateful to Giovanni Landi for useful conversation
about mathematical works on the subject.
F. F. and P. T. are grateful to CAPES for supporting their Ph.D.
projects. I. Sh. is grateful to CNPq, FAPEMIG and ICTP for partial
support. Also, I. Sh. is grateful to the Mainz Institute for
Theoretical Physics (MITP) for its hospitality and partial
support during the completion of the first version of this
work.
%

\begin{thebibliography}{m}

\bibitem{Polyakov} A. M. Polyakov,
Phys. Lett. B {\bf 103} (1981) 207.

\bibitem{Paneitz} S. Paneitz, 
MIT preprint, 1983;
SIGMA {\bf 4} (2008) 036.

\bibitem{rei} R. J. Riegert,
Phys. Lett. B {\bf 134} (1984) 56.
\\
E. S. Fradkin and A. A. Tseytlin,
Phys. Lett. B {\bf 134} (1984) 187.

\bibitem{BarrosShapiro} A. de Barros and I. L. Shapiro,
Phys. Lett. B \textbf{412} (1997) 242.

\bibitem{Conf1} C. R. Graham, R. Jenne, L. J. Mason and G. A. J. Sparling,
 J. London Math. Soc. {\bf s2-46 (3)} (1992) 557. 

\bibitem{Conf2} R. J. Baston and M. G. Eastwood,
\emph{Invariant operators, Twistors in mathematics and physics},
London Math. Soc. Lecture Notes {\bf 156}
(ed. T. Bailey and R. Baston, University Press, Cambridge, 1990).

\bibitem{Conf3} T. P. Branson,
Comm. Part. Diff. Equations {\bf 1} (1982) 393. 
Supp. Rend. Cir. Mat. Palermo, {\bf II 21} (1989) 115. 

\bibitem{Branson-85} T. P. Branson,
Math. Scand. {\bf 57} (1985) 293. 

\bibitem{Arak} T. Arakelyan, D. R. Karakhanyan, R. P. Manvelyan and R. L. Mkrtchyan,
Phys. Lett. B {\bf 353} (1995) 52.

\bibitem{Hamada} K. Hamada,
Prog. Theor. Phys. {\bf 105} (2001) 673.

\bibitem{Osborn} H. Osborn and A. Stergiou,
JHEP {\bf1504} (2015) 157.

\bibitem{Ansel} D. Anselmi,
Nucl. Phys. B {\bf 567} (2000) 331.

\bibitem{PoImpo} I. L. Shapiro,
Class. Quant. Grav. {\bf25} (2008) 103001.

\bibitem{MaMo} P. O. Mazur and E. Mottola,
Phys. Rev. D {\bf 64} (2001) 104022.

\bibitem{Stud}
D. F. Carneiro, E. A. Freitas, B. Gon\c{c}alves,
A. G. de Lima and I. L. Shapiro,
Grav. and Cosm. {\bf 40} (2004) 305.

\bibitem{capperkimber} D. M. Capper and D. Kimber,
J. Phys. A \textbf{13} (1980) 3671.

\bibitem{Weyl} G. de Berredo-Peixoto and I. L. Shapiro,
Phys. Rev. D {\bf70} (2004) 044024; \ 
Phys. Rev. D {\bf71} (2005) 064005.

\bibitem{ETG-76} F. Englert, C. Truffin and R. Gastmans,
Nucl. Phys. B {\bf 117} (1976) 407.

\bibitem{FrVi-78} E. S. Fradkin and G. A. Vilkovisky,
Phys. Lett. B {\bf73} (1978) 209. 

\bibitem{Grinshtein} B. Grinstein, A. Stergiou and D. Stone,
JHEP {\bf 1311} (2013) 195.

\bibitem{Solodukhin} A. F. Astaneh and S. N. Solodukhin,
arXiv:hep-th/1504.01653.

\bibitem{Intr} C. Cordova, T. T. Dumitrescu, K. Intriligator,
arXiv:hep-th/1506.03807.

\bibitem{DeserSchwimmer}
S. Deser and A. Schwimmer,
Phys. Lett. B {\bf 309} (1993) 279. 

\bibitem{DDI-76} S. Deser, M. J. Duff and C. J. Isham,
Nucl. Phys. B {\bf 111} (1976) 45.

\bibitem{decanini} Y. D\'ecanini and A. Folacci,
Class. Quant. Grav. {\bf 24} (2007) 4777. 

\bibitem{Myers} R. C. Myers and B. Robinson,
JHEP {\bf 1008} (2010) 067.

\bibitem{cadabra01} K. Peeters,
Comput. Phys. Commun. {\bf 176} (2007) 550. 

\bibitem{cadabra02} K. Peeters,
SPIN-06-46, ITP-UU-06-56,
arXiv:hep-th/0701238.

\bibitem{Elvang} 	
H. Elvang, D. Z. Freedman, L. -Y. Hung, M. Kiermaier, R. C. Myers
and S. Theisen,
JHEP {\bf 1210} (2012) 011.

\bibitem{FaSh} F. M. Ferreira and I. L. Shapiro, work in progress.

\bibitem{birdav} N. D. Birell and P. C. W. Davies,
\emph{Quantum Fields in Curved Space},
(Cambridge University Press, Cambridge, 1982).

\bibitem{duff94} M. J. Duff,
Class. Quant. Grav. {\bf 11} (1994) 1387. 

\bibitem{Deser-N} S. Deser,
Phys. Lett. B {\bf 479} (2000) 315; \ 
Helv. Phys. Acta {\bf 69} (1996) 570. 

\bibitem{Wunsch-86} V. W$\ddot{\rm u}$nsh,
Math. Nahr. {\bf 129} (1986) 269. 

\bibitem{GoverPeterson} A. R. Gover and L. J. Peterson,
Comm. Math. Phys. {\bf 235} (2003) 339. 

\bibitem{Waldron} M. Grigoriev and A. Waldron,	
Nucl. Phys. B {\bf 853} (2011) 291.

\bibitem{ParkerToms} L. Parker, D. Toms,
\emph {Quantum Field Theory in Curved Spacetime},
(Cambridge University Press, Cambridge, 2009).

\end{thebibliography}
%

\end{document}